\documentclass[twocolumn,floatfix,aps,prl]{revtex4-1}

\usepackage{hyperref}
\usepackage{graphicx}
\usepackage{caption}
\usepackage{amssymb,amsmath,amsthm,amsfonts}

\newcommand{\beq}{\begin{equation}}
\newcommand{\eeq}{\end{equation}} 

\begin{document}

\title{Supersolid behaviour of a dipolar Bose-Einstein condensate confined in a 
tube}
\author{Santo Maria Roccuzzo$^{1}$ and Francesco Ancilotto$^{2,3}$} 
\affiliation{$^1$INO-CNR BEC Center and Dipartimento di Fisica, Universit\`a di Trento,
238123 Povo, Italy
\\ $^2$Dipartimento di Fisica e Astronomia 
"Galileo Galilei" and CNISM, Universit\`a di Padova, 
Via Marzolo 8, 35122 Padova, Italy \\
$^3$CNR-IOM Democritos, via Bonomea, 265 - 34136 Trieste, Italy} 

\begin{abstract} 
Motivated by a recent experiment [L.Chomaz et al., Nature Physics {\bf 14}, 442 (2018)], 
we perform numerical simulations of a dipolar 
Bose-Einstein Condensate (BEC) in a tubular, periodic confinement
at T=0 within Density Functional Theory, where the
beyond-mean-field 
correction to the ground state energy is included in 
the Local Density Approximation.
We study the excitation spectrum of the system by solving 
the corresponding Bogoliubov-de Gennes equations. 
The calculated spectrum shows a 
roton minimum, and the roton gap decreases
by reducing the effective scattering length. 
As the roton gap disappears, the system spontaneously 
develops a periodic, linear
structure formed by denser clusters of atomic dipoles 
immersed in a dilute superfluid background. 
This structure shows the hallmarks of a supersolid system, 
i.e. (i) a finite non-classical translational inertia 
along the tube axis and (ii) the appearance of two gapless modes, i.e. a phonon mode
associated to density fluctuations and resulting from  
the translational discrete symmetry of the system, and the  
Nambu-Goldstone gapless mode corresponding to phase fluctuations,
resulting from the spontaneous breaking of the gauge symmetry.
A further decrease in the scattering length eventually leads to the
formation of a periodic linear array of self-bound droplets.
\end{abstract} 

\date{\today}
\maketitle
Dipolar Bose Einstein condensates (BECs) 
attracted great attentions in recent years, 
since the first experimental realizations of
BECs with strongly magnetic atomic gases\cite{bec_chromium,bec_dy,bec_er}.
This interest is motivated by the particular properties of such systems which 
are characterized by anisotropic and long-range dipole-dipole interactions
in addition to the short-range contact interactions,
resulting in a geometry dependent stability diagram \cite{Koch2008}
where the system (which is intrinsically unstable in 
3D) becomes stable against collapse if the
confinement along the polarization axis is much tighter that the
in-plane confinement. The properties of dipolar BECs have been
the subject of numerous experimental and theoretical studies, extensively reviewed
in Ref.\cite{lahaye,baranov}.

Recent experiments \cite{Kadau2016,onset} on the stability of a dipolar BEC 
of $^{164}$Dy trapped in a flat "pancake" trap
showed the formation of droplets arranged in an ordered 
structure, their collapse being prevented by the tight confinement
along the short axis.
This effect is the equivalent of the Rosensweig instability of classical ferrofluids
\cite{rosensweig67}. 

Remarkably, recent experiments\cite{schmitt} showed that 
{\it self-bound} droplets can be realized in a dipolar Bose gas  
depending upon the ratio between the strenghts 
of the long-range dipolar attraction 
and the short range contact repulsion.
These droplets, whose densities are higher by about one order of
magnitude than the density of the weakly interacting condensate, 
are stable even in free space, after the external trapping potential 
is removed.

The possibility of self-bound dipolar droplets 
has been explained theoretically in Ref.\cite{santos_wachtler,blackie_baillie1,ferlaino}, 
where it has been shown that 
the binding arises from the interplay between the 
two-body dipolar interactions and the effects of quantum fluctuations.
The latter can be embodied in a beyond-mean-field energy correction
\cite{lima_pelster,santos_wachtler},
where a positive shift of the ground state energy 
with the Lee-Huang-Yang (LHY) form \cite{hy1957} counteracts 
the destabilizing effect of the dipole-dipole attraction.
The crossover in a dipolar quantum fluid from a 
dilute BEC to self-bound
macrodroplets was studied in Ref.\cite{ferlaino}, where 
further evidence was provided that quantum fluctuations indeed stabilize the
ultracold gas far beyond the instability threshold imposed by mean-field interactions.
The properties 
of self-bound dipolar quantum droplets have been 
extensively studied from a computational point of 
view, both within a mean field theory approach that
takes into account the LHY 
correction \cite{blackie_baillie1,blackie_baillie2}, and with
Quantum Monte Carlo simulations \cite{saito,boronat,salas}. 

In Ref. \cite{Chomaz2018} it has been 
shown that in a dipolar BEC of $^{166}$Er confined 
in a strongly prolate cygar-shaped trap ("tubular" trap), 
the reduction of the scattering length leads 
to the appearance of a roton mode.
The excited states dispersion relation is thus 
characterized by a roton minimum, similarly to
the case of $^4$He, the roton gap amplitude depending
on the relative strengths of short-range and dipolar interactions,
as predicted in Ref.\cite{lewe,odell}.
This suggests that when the roton gap becomes very small,
a dipolar BEC confined in an axially elongated trap orthogonal 
to the polarization direction may develop a modulated
density profile in its ground state. Based on this, it
has been suggested\cite{Chomaz2018}  
that this system may indeed show supersolid behavior.

The existence of a supersolid phase of matter was proposed long ago
for $^4$He \cite{gross_sup}, but its experimental verification remained
elusive \cite{kim2012}.
The possibility of forming a solid structure simultaneously 
possessing crystalline order and superfluid properties\cite{boninsegni}
is associated with an excitation spectrum 
of the liquid phase characterized by a roton minimum 
at finite k-vector\cite{kirz}, the liquid to ”supersolid” transition 
being triggered by the vanishing of the roton gap.
Supersolid phases have been recently predicted for confined 
condensed spinless bosons in 2-dimensions\cite{cinti} and 3-dimensions\cite{ancilotto}
interacting via a broad class of soft core repulsive potentials.

Supersolid behavior has been proposed for 
the stripe phase of a dipolar Bose gas under strong confinement 
when the polarization axis forms an angle with 
the tight confinement axis\cite{bombin,cinti1,wenzel}. 
Similar predictions have been made for a dipolar BEC confined
in a quasi-2D pancake shaped trap \cite{blackie_baillie3}, where a possible
supersolid behaviour is related to the formation of a low-density "halo" of atoms
among different droplets in a cluster arrangement when the chemical potential
is high enough to let some atom escape from a single droplet. 
Finally,
a supersolid behavior has been suggested in a ferrofluid mixture of
dipolar BEC under a "pancake" confinement\cite{saito2}, within
a mean-field approach.

The only experimental evidence so far of supersolid behavior in 
cold gases has been reported recently in Ref.\cite{leonard},
where the authors realized an "infinitely stiff" supersolid of $^{87}Rb$ atoms
with the density modulation artificially imposed by 
external optical lattices.
Stable "stripes" modulations have been experimentally observed
recently in dipolar quantum gas\cite{wenzel,tanzi}.
While no global phase coherence is found in a similar system 
studied in Ref.\cite{wenzel},
a partial phase coherence is suggested in Ref.\cite{tanzi},
thus indicating possible supersolid beavior.

We notice that in the systems studied in Ref.\cite{blackie_baillie3,wenzel}
the condensate $\rightarrow$ droplet transition  
results in the formation of finite clusters made of few "stripes"
(i.e. very elongated droplet in the polarization direction).
In both of the above two studies   
many local minima of the total energy are possible 
depending upon the number of atoms in the condensate, and these minima
are characterized by different number/arrangement of droplets.
One of such state appears to be a stationary state with
global coherence that is predicted to be robust against quantum phase
fluctuations\cite{blackie_baillie3}.

We will propose in the following a different geometry,
%which allow to avoid this multi-configuration scenario,
where the condensate-droplet transition occurs 
in a tubular confinement with periodic boundary conditions,
resulting in a density modulated 
configuration made by a linear, periodic arrangement
of equally spaced elongated "droplets" immersed in a halo of 
low-density superfluid. We will provide here evidence
of the supersolid character of such structure.

In this paper we will use
numerical simulations within
Density Functional theory (DFT) at T=0,
in the Local Density approximation (LDA), to study
the equilibrium structure and elementary 
excitations of a dipolar BEC confined in a tube whose
axis is orthogonal to the polarization direction,
and with periodic boundary conditions along the tube axis.
%This choice makes the studied system a model for 
%possible experimental realization employing a ring 
%trapping potential, as will be explained in the following.

%We will show that 
%the transition from the stable BEC to an ordered lattice 
%of quantum droplets can be triggered
%by the softening of the roton gap,
%which is in turn driven by a reduction in the scattering
%length, resulting in an increase of the ratio
%between the strengths of the dipole-dipole and contact
%interactions.
%As the gap disappears, a finite 
%fraction of the atoms is detached from the droplets, and so
%can form a "halo" of dilute superfluid in which the 
%system is immersed, eventually realizing a supersolid phase.
%For higher values of the interaction parameter 
%For higher values of the ratio between the strengths of the dipole-dipole
%and contact interactions 
%the superfluid structure turns into an ordered array 
%of self-bound droplets, 
%where the supersolid behaviour is suppressed.

Within the DFT framework, the total energy of
a dipolar BEC of atoms with 
mass $m$ and magnetic moment ${\bf\mu}$ is:

\begin{widetext}
\begin{eqnarray}
E=\int \Big [{\hbar ^2 \over 2m} |\nabla 
\phi ({\bf r})|^2+V_t({\bf r})|\phi ({\bf r})|^2
+{g\over 2}|\phi ({\bf r})|^4
\Big ]d{\bf r}+
{1\over 2}{\int V_{dd}(|{\bf r}-
{\bf r}^\prime |)|\phi ({\bf r})|^2  
|\phi ({\bf r}^\prime )|^2d{\bf r}\,d{\bf r}^\prime}
+{2\over 5}\gamma (\epsilon _{dd})\int |\phi ({\bf r})|^5d{\bf r}
\end{eqnarray}
\label{energy}
\end{widetext}
Here $g=\frac{4\pi\hbar^2a}{m}$, $a$ being the s-wave scattering length, 
$V_{dd}({\bf r}-{\bf r'})=\frac{\mu_0\mu^2}{4\pi}\frac{1-3\cos^2\theta}{|{\bf r-{\bf r'}}|^3}$
is the dipole-dipole interaction between two identical magnetic dipoles 
aligned along the z axis 
($\theta $ being the
angle between the vector ${\bf r}$ and the polarization direction $z$), and
$\mu_0$ is the permeability of the vacuum.
$V_t$ is the trapping potential.
The last term is the beyond-mean-field (Lee-Huang-Yang, LHY) correction \cite{lima_pelster},
where $\gamma(\epsilon_{dd})=
\frac{32}{3\sqrt{\pi}}ga^{\frac{3}{2}}F(\epsilon_{dd})$, 
$\epsilon_{dd}=\frac{\mu_0\mu^2}{3g}$ 
being the ratio between the strenghts of the dipole-dipole and
contact interactions, and 
$F(\epsilon_{dd})=\frac{1}{2}\int_0^{\pi}d\theta 
\sin\theta[1+\epsilon_{dd}(3\cos^2\theta-1)]^{\frac{5}{2}}$. 
%In the case of polar molecules the same expression for $V_{dd}$ applies, 
%with $\mu _0 \mu _d^2$ replaced by 
%$d^2/\epsilon _0$, where $d$ is the electric dipole moment
%of the molecule and $\epsilon _0$ is the vacuum permittivity.
The number density of the dipole system is $n({\bf r})=|\phi ({\bf r})|^2$.

The minimization of the above energy functional
leads to the following Euler-Lagrange equation:
%\begin{align}
% H_0 \phi({\bf r})=\mu  \phi({\bf r})
% \label{acca}
%\end{align}
%with
\begin{align}
 H_0\phi({\bf r})\equiv
& \big[-\frac{\hbar^2}{2m}\nabla^2+V_t({\bf r})+g|\phi({\bf r})|^2+
\gamma(\epsilon_{dd})|\phi({\bf r})|^3+
\nonumber\\
  & \int d{\bf r'}|\phi({\bf r'})|^2V_{dd}({\bf r}-{\bf r'})
   \big]\phi({\bf r})=\mu  \phi({\bf r})
 \label{acca}
\end{align}
and
$\mu $ is a Lagrange multiplier whose value is determined
by the normalization condition $\int |\phi ({\bf r})|^2 d{\bf r}=N$
($N$ being the total number of dipoles).
Eq.(\ref{acca}) is the 
well-known Gross-Pitaevskii equation \cite{gp} with
the addition of the LHY correction.
In what follows $m$ is the mass of a $^{166}$Er atom.
A similar approach 
has been used, e.g., in Ref.
\cite{santos_wachtler} and other papers addressing the 
effect of beyond-mean-field effects on the dipolar Bose gas.
The predictions of the DFT-LHY approach described above
has been tested 
in Ref.\cite{saito} against Quantum Monte Carlo simulations,
showing that the DFT-LHY indeed allows rather accurate predictions.

In the following we will 
assume a tubular confinement, i.e. 
the dipoles are radially confined by an harmonic potential
$V_t({\bf r})=\frac{1}{2}m(\omega^2_yy^2+\omega^2_zz^2)$,
in the y-z plane (z is the polarization direction and y is the transvers
direction).
The harmonic frequencies 
are fixed to the values $\omega_y=\omega_z=2\pi(600)Hz$.
This geometry closely matches the experimental set-up used in 
the recent experiments of Ref.\cite{Chomaz2018,ferlaino}.
Along the third axis, $x$, the system is not confined, but subject
to periodic boundary conditions (PBC), $\phi(x+L,y,z)=\phi(x,y,z)$, 
$L$ being the tube length. 
Note that, due to the presence of PBC, 
the system is equivalent to a ring geometry 
(with a ring radius $R=L/2\pi$),
if curvature effects can be neglected (i.e. when $R$ is much larger that the 
harmonic confinement length in the y-z plane).
This allows to test our prediction in actual experiments,
where ring-shaped trapping potential can be easily realized. 

We solve equation (\ref{acca})
by propagating in imaginary time its time-dependent 
counterpart $i\hbar \partial \phi/\partial t=H_0\phi$.
In all the simulations we 
fix the value of the linear density $n_0=N/L$ and vary the value of
the ratio, $\epsilon_{dd}$, between the dipolar and contact interaction strengths. 
The total number of atoms is fixed to $N=6\times 10^4$.
To compute the spatial derivatives
appearing in the (\ref{acca}),
we used an accurate 13-point finite-difference formula.
Density $n$ and wavefunction $\phi $ are represented in 
real space on a three-dimensional spatial mesh with 
%SANTO: completa
spacing $h=0.1\mu$m.
The convolution integral in the potential energy term
of Eq.(\ref{acca}) is efficiently evaluated 
in reciprocal space by using Fast Fourier transforms,
recalling that the Fourier transform
of the dipolar interaction is \cite{lahaye}
${\tilde V}_{\bf k}=(\mu _0 \mu/3)(3\cos^2\alpha -1)$
where $\alpha$ is the angle between ${\bf k}$ and
the $z$-axis. We verified that the transverse dimensions
of our simulation cell are wide enough to make 
negligible the effects, on the energy values and density profiles,
of the spurious dipole-dipole interaction 
between periodically repeated images.

In order to study the elementary excitations, we 
expand the wave function in the Bogoliubov-de Gennes (BdG) form 
$\Phi({\bf r},t)=e^{-i\frac{\mu}{\hbar}t}[\phi({\bf r})+
u({\bf r})e^{-i\omega t}-v^*({\bf r})e^{i\omega t}]$, and
insert this expansion in Eq.(\ref{acca}). 
Keeping only terms linear in the amplitudes $u$ and $v$, 
one gets the BdG equations for the amplitudes $u$ 
and $v$ and the excitation energies $\epsilon $, that can be 
cast in a matrix form as\cite{blackie_baillie2}:

\beq
 \label{bdgeq}
 \begin{pmatrix}
  H_0-\mu+\hat{X} && -\hat{X}^{\dagger} \\
  \hat{X}     && -H_0+\mu +\hat{X}^{\dagger}
 \end{pmatrix}
 \begin{pmatrix}
  u \\
  v
 \end{pmatrix}
 =\epsilon
 \begin{pmatrix}
  u\\
  v
 \end{pmatrix}
\eeq
where $H_0$ is given in Eq.(\ref{acca})
and the operator $\hat{X}$ is defined by its action on the function $f$ as
\begin{widetext}
\begin{align}
 \hat{X}f({\bf r})= & \phi({\bf r})\int d{\bf r'}
[V_{dd}({\bf r}-{\bf r'})+g\delta({\bf r}-{\bf r'})]\phi^*({\bf r'})f({\bf r'})+
      \frac{3}{2}\gamma(\epsilon_{dd})|\phi({\bf r})|^3f({\bf r}) 
\end{align}
\end{widetext}

Because of our use of Fourier transforms,
which imply that PBC must be imposed in our calculations,
we can expand the wavefunction $\phi $ and
the complex functions $u,v$ in the Bloch form
appropriate to a periodic system. In this way, the 
equations (\ref{bdgeq}) can be solved in reciprocal space
allowing to find $\epsilon _{\bf k}$ in 
the right-hand side of Eq.(\ref{bdgeq})
(see Ref.\cite{ancilotto} for details about the numerical methods 
used to solve Eq.(\ref{bdgeq})).

\begin{figure}
 \includegraphics[width=\columnwidth]{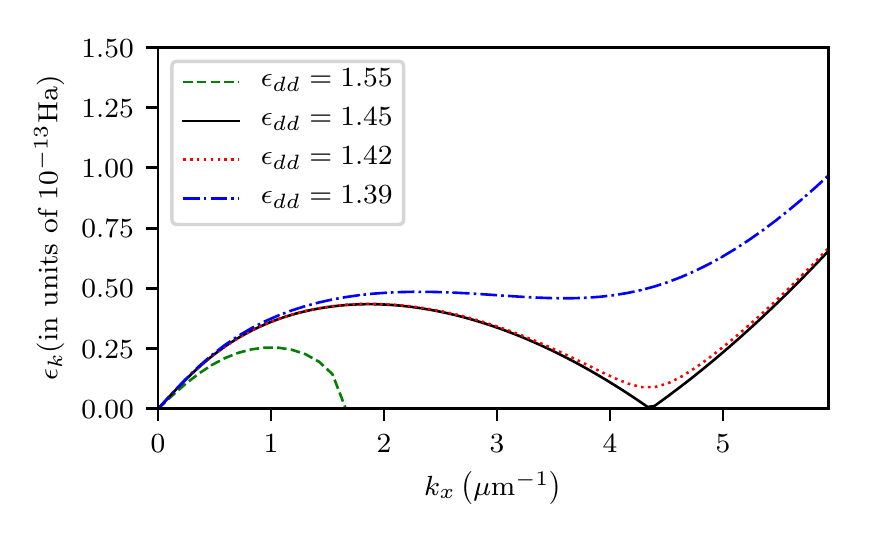}
 \includegraphics[width=\columnwidth]{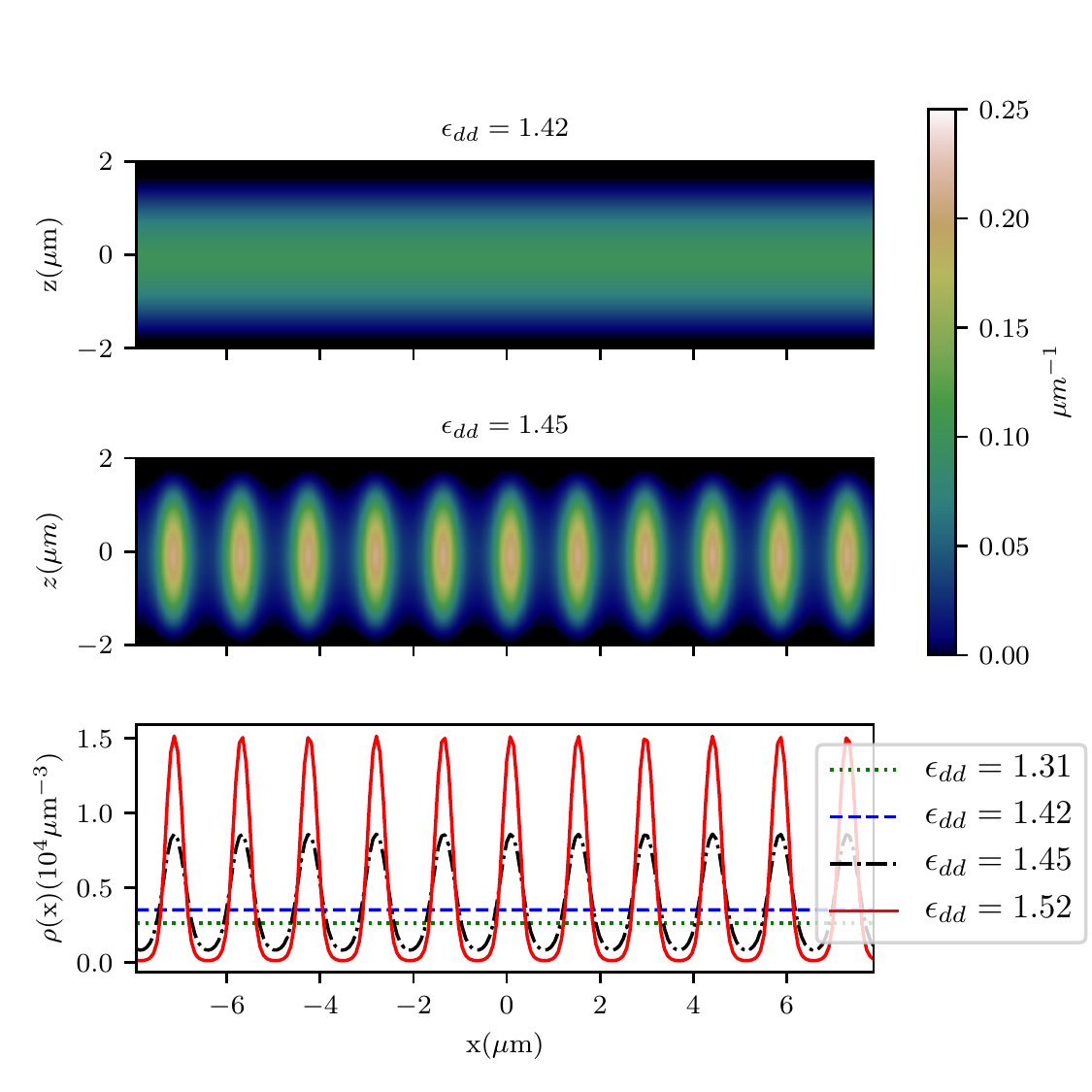}
 \caption{Upper panel:
          dispersion relation of excitations propagating along the tube axis
in the homogeneous system. Energies are in atomic units.
            Lower panel: integrated density $n_y(x,z)$ just 
below and at the 
            the critical value of $\epsilon _{dd}$ where the
roton gap vanishes. The total number of atoms is $N=6\times 10^4$.
The lowest plot shows the density $n$ along the tube axis for different
values of $\epsilon _{dd}$. 
                    }
 \label{int_den}
\end{figure}

We first solve the BdG equations to compute the excitation 
spectrum for a dipole system 
characterized by a uniform density along the 
tube axis (x-axis). The energies $\epsilon _{\bf k}$ 
of the mode along the $k_x$ direction are shown in 
figure \ref{int_den} (upper panel) for the choice $n_0=3.78\times10^3\mu\mbox{m}^{-1}$, 
for different values of 
$\epsilon_{dd}$. Notice that, 
as $\epsilon_{dd}$ is increased (i.e. the scattering length 
$a$ is decreased), a roton minimum develops
in the dispersion relation, 
eventually vanishing at $\epsilon_{dd}=1.45$.

This signals a possible density modulation instability that might 
break the uniform symmetry along the tube axis.
In order to verify this, we calculated the equilibrium 
density profile by solving the Eq.(\ref{acca}) for different valus of $\epsilon_{dd}$.
In figure (\ref{int_den}) we show the resulting density 
for two different values of $\epsilon_{dd}$.
%As the typical imaging techniques in experiments allows
%to measure the density of the BEC integrated along the viewing direction,
We plot in figure (\ref{int_den}) the density $n_y(x,z)=\int n(x,y,z)dy$ 
integrated along the y-axis perpendicular to the polarization
direction. One can see
that the density remains uniform 
along the tube for finite values of the roton gap,
while it becomes periodically modulated as the roton gap 
vanishes. The resulting structure in the latter case is 
shown in the lower panel of figure (\ref{int_den}).
The periodicity of the density profile is fixed by
$\lambda=\frac{2\pi}{k^c_x}$, where $k^c_x$ is the critical
value of the momentum at which the roton gap vanishes.
When the tube length is not commensurate with the roton 
wavelength, as is the case shown in the figure, the
modulation develops at a wavelength most close to it.
Such periodic modulation is maintained 
well below the transition,
as shown in the lowest plot in figure (1).
%We notice that the above 
%structure is robust, i.e.  
%the imaginary-time evolution always converge to this state 
%irrespective of the perturbation that we add
%to the uniform initial state (in the form of random noise
%with different amplitudes).

If we start instead from an initial state modulated 
with a wavelength different from $2\pi/k^c_x$,
we sometimes got trapped, during the minimization procedure,
into metastable states characterized by a 
different number of stripes, with a higher energy than the 
state shown in Fig.1.
This happens, for instance, with a state having 12 or 9
stripes in the tube (for values of $\epsilon_{dd}$ close to the roton 
instability value $\epsilon_{dd}=1.45$), instead of the 11 stripes found 
for the ground-state (a 10-stripes solution is found to be 
unstable towards the lowest-energy 11-stripes
structure, i.e. it always evolves towards it
during the imaginary-time evolution).
The energy differences with respect to 
the ground-state are however very small (the 12 stripes state being almost degenerate
with the 11 stripes one, with just a 0.1\% relative difference, while 
we find a 1\% relative difference for the 9 stripes case). 
This implies that during a rapid quench of the 
interaction the system might indeed get caught into one of such metastable
states. The resulting structures, however, still
have supersolid properties, being characterized
simultaneously by periodic order and a finite non-classical translational inertia
(see the following).
Finally, although close to the roton instability our structure is 
a bona-fide ground-state, we cannot exclude that for higher 
values of $\epsilon_{dd}$ the solution we find is a 
metastable state rather than the ground-state. 

The periodic structure corresponding to $\epsilon_{dd}=1.45$
appears to be made of regularly arranged, dense elongated 
clusters of dipoles immersed
in a background of very dilute condensate, 
as shown in the lowest density plot of figure (\ref{int_den}).
This suggests that the systems, for $\epsilon_{dd}>1.45$
may display a supersolid character. In order
to verify this hypothesis, we have looked for 
the characteristic hallmarks of supersolid behavior
of the modulated structures, i.e.\cite{sep_joss_rica}
(i) a finite non-classical translational inertia 
and (ii) the appearance, besides the phonon mode associated with the 
density periodicity, 
of a gapless "superfluid band" resulting 
from the spontaneous breaking of global gauge symmetry.

First, we check
the presence of Non-Classical Translational Inertia (NCTI).
This is done by solving for stationary states the 
real-time version of the EL equation (\ref{acca}) in the 
comoving reference frame with uniform velocity $v_x$, i.e.
\begin{equation}
 i\hbar\frac{\partial}{\partial t}\phi({\bf r})=  
(H_0+i\hbar v_x\frac{\partial}{\partial x})\phi({\bf r})
\end{equation}

Following Ref. \cite{sep_joss_rica}, we define the superfluid fraction $f_s$, 
as the fraction of particles that 
remains at rest in the comoving frame:
\beq
 f_s=1-\lim_{v_x\to 0} \frac{{\big <}P_x{\big >}}{Nmv_x}
\eeq 
where ${\big <}P_x{\big >}=-i\hbar \int \phi ^{\ast} \partial \phi /\partial x$ 
is the expectation value of the momentum and $Nmv_x$ is 
the total momentum of the system if all the particles 
were moving
($f_s$ should not confused with the
total superfluid fraction: for instance 
in the deep non-linear regime where self-bound droplets form, as
shown in the following,
although they are individually
superfluid, $f_s$ zero, 
meaning that there is no supersolid behavior).

The definition above is the most natural to reveal global phase 
coherence in a periodic system like the one studied here\cite{sep_joss_rica}.
Other ways of quantifying the tunneling 
in strongly confined systems made of a cluster of droplets 
are possible, like e.g. 
approximately treating pairs of droplets as 
bosonic Josephson junctions\cite{wenzel,blackie_baillie3}.

\begin{figure}
 \includegraphics[width=\columnwidth]{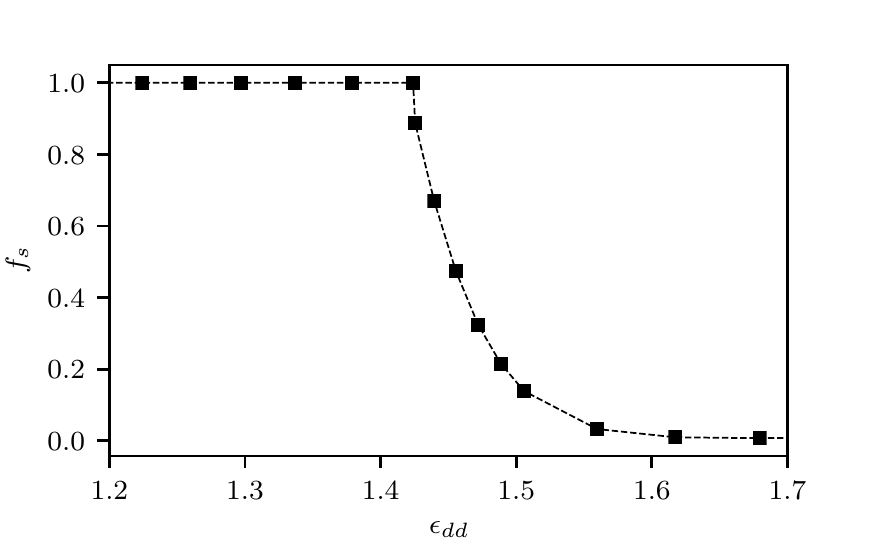}
 \includegraphics[width=\columnwidth]{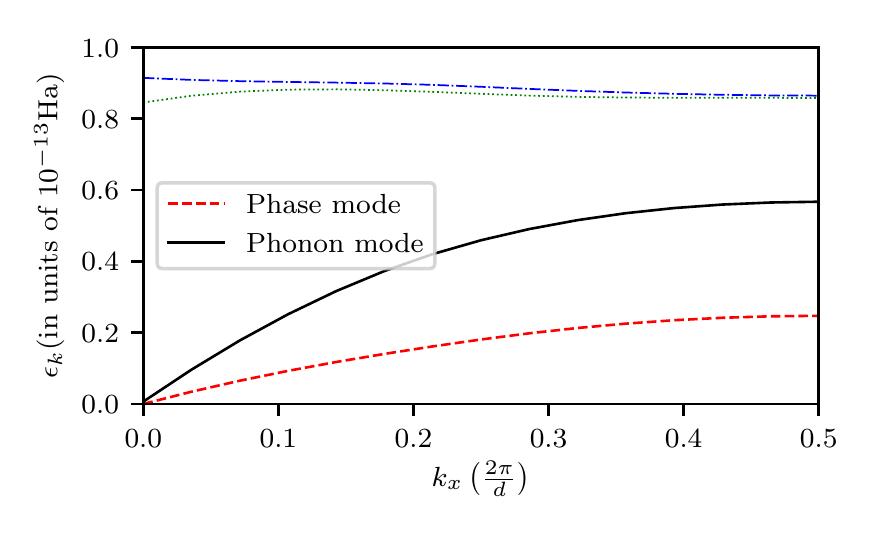}
 \caption{Upper panel:
           superfluid fraction as function of $\epsilon_{dd}$. Lower panel:
            excitation spectrum along the tube axis, calculated for 
%SANTO: completa
$\epsilon_{dd}=1.45$. The rightmost value of 
           $k_x$ corresponds to the $1^{st}$ Brillouin zone boundary along the 
%SANTO: controlla che quanto detto qui di seguito sia corretto 
x-axis, i.e. $k_x=\pi/d$, $d=L/11$ being the length of the unit cell containing
exactly one droplet in Fig.(\ref{int_den}).
             }
 \label{supfrac}
\end{figure}

We can see from figure (\ref{supfrac})
that, when a modulation in the density profile appears, 
the superfluid fraction becomes smaller than one,
and it decreases as $\epsilon_{dd}$ is increased. 
A small jump at the uniform $\rightarrow$ modulated transition 
seems to signal a first-order transition,
similarly to what found in the case of supersolid transition
of soft-core bosons\cite{ancilotto}. 

Another characteristic of supersolid behavior is associated with the presence
in the excitation spectrum
of the periodically modulated structure shown in 
figure (\ref{int_den}), of an extra gapless mode besides the 
"phonon" modes associated with the periodic density modulations\cite{nambu}. 
The excitation spectrum can be calculated by 
solving the corresponding BdG equations for modes 
propagating along the axis of the tube. 
The result is shown in figure (\ref{supfrac}), 
for the values $\epsilon _{dd}=1.45$ and $n_0=3.78\times10^3\mu\mbox{m}^{-1}$,
from which one can see the appereance of two gapless
modes associated with symmetry breaking.
The harder mode 
is associated to the density response of the system, 
and it corresponds to the phonon branch. 
The softer mode is associated instead to the 
phase response of the system, and it signals 
the superfluid character of the supersolid
(Nambu-Goldstone mode).
The correct mode assignement was made by 
looking at the calculated local density and phase fluctuation 
modes \cite{macri,ancilotto}, 
$ \Delta\rho _{n{\bf k}}({\bf r})= 
|u_{n,{\bf k}}-v_{n,{\bf k}}|^2$ and
$\Delta\theta _{n{\bf k}}({\bf r})= |u_{n,{\bf k}}+v_{n,{\bf k}}|^2$, respectively:
the phonon mode is mainly characterized by density modulations, whereas
the superfluid mode is characterized mainly by modulations in the 
phase.
As $\epsilon _{dd}$ increases, the system is entering the regime 
of self-bound droplets (as discussed below), and as a result the
Goldstone phase mode become softer and softer, until it completely disappear.
In this regime the droplets are disconnected from one
another and the superfluid fraction associated with the
non-classical inertia goes to zero, while the individual droplets are
still superfluid.

%To identify unambiguously the two modes, we can also 
%calculate local density and phase fluctuations
%following Ref. \cite{macri}, that is as
%\begin{align}
%& \Delta\rho({\bf r})={\big<}\delta\rho^{\dagger}({\bf r})
%\delta\rho({\bf r}){\big>}/|\phi({\bf r})|^2=\sum_{n,{\bf k}}
%|u_{n,{\bf k}}-v_{n,{\bf k}}|^2\nonumber\\
%&\Delta\theta({\bf r})={\big<}\delta\phi^{\dagger}({\bf r})
%\delta\phi({\bf r}){\big>}\times4|\phi({\bf r})|^2=
%\sum_{n,{\bf k}}|u_{n,{\bf k}}+v_{n,{\bf k}}|^2
%\end{align}

%In this way, we can isolate contributions to density 
%and phase fluctuations from single bands. In particular,
%isolating these contributions from the two bands 
%reported in Fig.(\ref{supfrac}) for $k_x=0.5\frac{\pi}{L}$,
%we find, as reported in figure (\ref{deltas}),
%that the second band gives a stronger contribution
%to density fluctuation then the first one, which in turn
%gives a major contribution to phase fluctuations. 

\begin{figure}
 \includegraphics[width=\columnwidth]{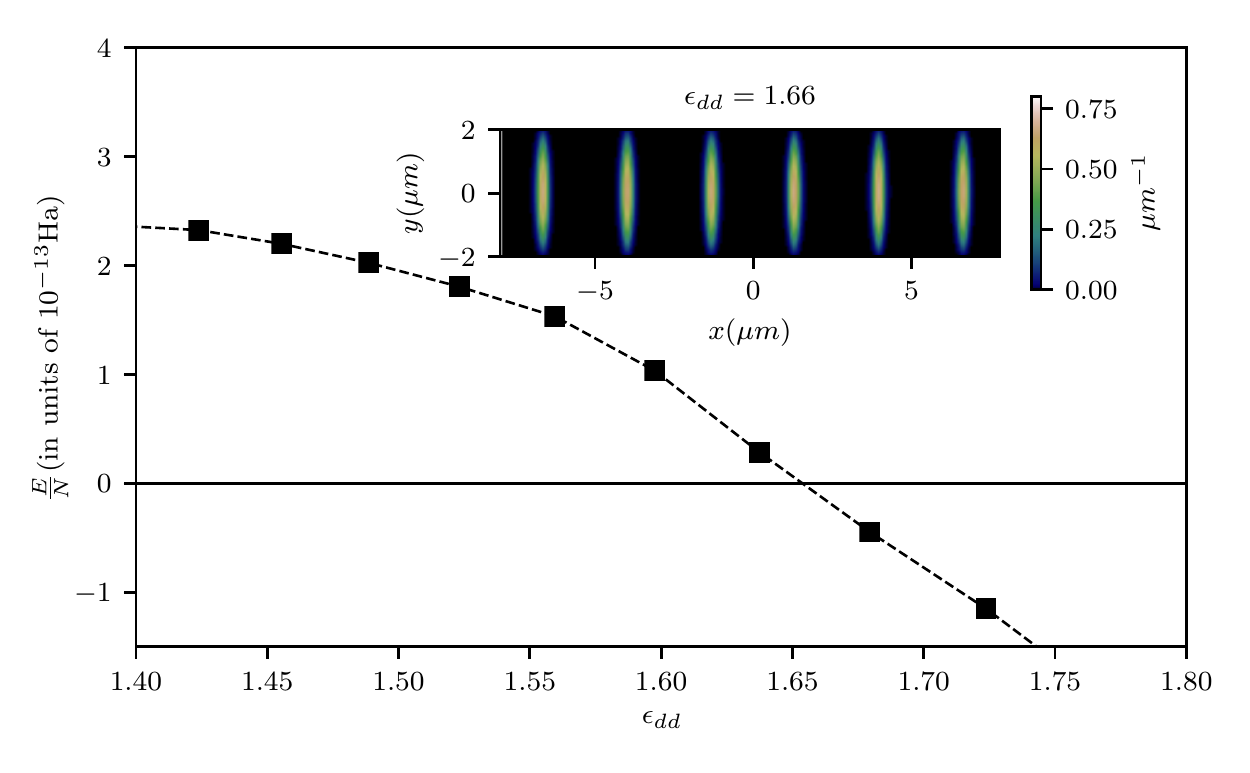}
 \caption{Energy per particle (in atomic units) as a function 
of $\epsilon _{dd}$. The inset shows the array of self-bound droplets.}
 \label{nrg}
\end{figure}

From figure (\ref{supfrac}) (upper panel) 
it appears that as $\epsilon _{dd}$ increases,
the superfluid fraction tends to zero. When this happens, the 
atomic clusters shown in the lowest panel of figure (\ref{int_den})  
begin to merge, forming denser isolated droplets, while
the calculated energy per particle 
becomes negative, as shown in figure (\ref{nrg}).
This happens at $\epsilon _{dd}\sim 1.66$.
Above this value, the density profile of the 
system takes the form of an (ordered) one-dimensional lattice of 
{\it self-bound} quantum droplets (each containing $N\sim 10^4$ atoms), 
while the supersolid 
behaviour is completely suppressed, as shown by the disappearance
of the Nambu-Goldstone mode from the 
the calculated excitation spectrum.

We notice at this point that we obtained similar results  
with different choices for the system density and tube length.
However, there is no special choice for such parameters
which will give supersolid properties. Rather,
for a given density, tube length and radial confinement,
there is always a range of coupling strengths where the
system shows supersolid behavior: a different choice of the 
parameters will only shift the condensate-supersolid transitions towards
different values of the coupling strength, but the
relevant physics will not be affected. 
(the density must however be large enough for the system to develop
the expected modulation).
We notice however that for very long tubes (much longer than 
the ones investigated here)
quantum phase fluctuations may
destroy the phase relationship between distant 
droplets.

%The inverse compressibility of the system 
%$ \kappa ^{-1}=V\frac{\partial^2E}{\partial V^2}$ can
%be easily computed by static calculations 
%with a fixed number of particle $N$, where
%the length of the tube is varied while the 
%transverse confinement remains unchanged.
%In figure (\ref{inc}) we show  
%$\kappa ^{-1} $ as a function of $\epsilon_{dd}$.
%One can see that, whereas the supersolid phase 
%turns out to be extremely compressible,
%($ \kappa ^{-1}\sim 0.03$
%for $\epsilon_{dd}=1.6$, in
%the units of Fig.(\ref{inc})), 
%as the supersolid behaviour is suppressed 
%$ \kappa ^{-1}$ 
%increases rapidly, i.e. 
%the compressibility of the system rapidly decreases. This is due 
%to the repulsive interaction between the self-bound droplets 
%which behave as giant aligned dipoles repelling each other.

%\begin{figure}
%\includegraphics[width=\columnwidth]{ic.pdf}
% \caption{Inverse compressibility (in atomic units) as a function of $\epsilon _{dd}$}
% \label{inc}
%\end{figure}

Preliminary calculations show that
the super-solid character of the system described here
is robust against weak perturbations of the external potential.
In particular, small periodic modulation of the trapping potential
do not destroy the supersolid behavior\cite{roccuzzo}.

In conclusions we have shown, 
by means of numerical simulatioons based on the DFT-LHY approach,
that in a dipolar BEC confined in a tube at $T=0$ 
the softening of the roton mode, caused by a decrease in the 
scattering length,
leads to the formation of a modulated, periodic structure,
in which denser clusters of
dipoles are immersed in a very dilute superfluid background.
This system shows the hallmarks of supersolid behaviour, i.e.
a finite, non-classical translational inertia, and
a Goldstone "superfluid" mode in the excitation spectrum in addition
to the phonon mode associated with density periodicity.
The supersolid behaviour is suppressed when the system, by further decreasing
the scattering length,
enters in a regime in which the dipole clusters turn 
into an ordered array of self-bound quantum droplets.
The tubular confinement is more convenient 
from the experimental point of view than a 2D confinement
because it likely reduces the number of possible metastable states
with comparable energies.

%Our results suggests that a possible experimental
%realization of a supersolid can be achieved 
%by confining a dipolar BEC in a ring trap, and by
%fine tuning the scattering length of the 
%system in order to increase the parameter $\epsilon_{dd}$
%across the superfluid-supersolid phase transition, coupling 
%the atoms with an external 
%magnetic field and using the knowledge of the Feshbach 
%resonances of the atomic species trapped. 

%At variance with the "infinite-stiffness" supersolid found in Ref. \onlinecite{leonard}, 
%where the density modulation is imposed by external fields,
%our finding suggest the possibility of realizing 
%a supersolid with a finite stiffness, where the compressibility 
%can be varied by tuning the interparticle interaction strength.

The phase coherence of the supersolid phase described here
could be experimentally 
detected in a dipolar condensate confined in 
a ring trap where, after having tuned the scattering
length across the threshold value required for 
the supersolid formation, 
the trapping potential is switched off,
allowing the system to expand freely.
By doing subsequent absorption imaging one could detect the 
presence of interference maxima associated with 
phase coherence\cite{leonard}. 

{\it Note added}
During the review process of the present manuscript
a joint experimental-theoretical paper appeared \cite{arx}
showing that a ground-state, coherent linear array of quantum "droplet" 
can be realized where, in addition to periodic density 
modulation, a robust phase coherence across the system is mantained,
similarly to what we predicted here.

\begin{acknowledgments}
We thank Alessio Recati, Luca Salasnich and Sandro Stringari for useful exchanges.
\end{acknowledgments}

\end{document}